\documentclass[fleqn,10pt]{wlscirep}
\usepackage[utf8]{inputenc}
\usepackage[T1]{fontenc}
\title{Prediction of typhoon tracks using a generative adversarial network with observational and meteorological data}

\author[1]{Mario Rüttgers}
\author[1]{Sangseung Lee}
\author[1,*]{Donghyun You}
\affil[1]{Pohang University of Science and Technology, Mechanical Engineering, Pohang, 37673, South-Korea}

\affil[*]{dhyou@postech.ac.kr}

\begin{abstract}

Tracks of typhoons are predicted using a generative adversarial network (GAN) with observational data in form of satellite images and meteorological data from a reanalysis database. Time series of images of typhoons which occurred in the Korean Peninsula in the past are used to train the neural network. The trained GAN is employed to produce a 6-hour-advance track of a typhoon for which the GAN was not trained. The predicted image favorably identifies the future location of the typhoon center as well as the deformed cloud structures. The errors between predicted and real typhoon centers are measured quantitatively in kilometers. $65.5~\%$ of all typhoon center predictions have an error of less than $80~km$, $31.5~\%$ lie within a range of $80 - 120~km$ and the remaining $3.0~\%$ are above $120~km$. The overall error is $67.2~km$, compared to $95.6~km$ when only observational data are used as input. The cloud structure prediction is evaluated qualitatively. It is shown that the GAN is able to predict trends in cloud motion. It is found that adding physically meaningful meteorological data to satellite images improves the sharpness of predicted images.

\end{abstract}
\begin{document}

\flushbottom
\maketitle
%
%
\thispagestyle{empty}

\section{Introduction}

In the past, the Korean Peninsula suffered from a multitude of typhoons. The destructive force of a typhoon comes from the wind speed of the rotating air and the rainfall that can cause disastrous flooding \cite{spiegel18}. The highest wind speeds are usually found near the typhoon center. Strong rain, on the other hand, can be experienced near dense cloud structures. This work focuses on predicting both, typhoon center coordinates and cloud motion. Together, both types of forecasts are defined as typhoon track predictions. Predictions are done by using observational and meteorological data as input to a deep learning method. The geographical focus lies on typhoons that affect the Korean peninsula.

Recently, various applications of neural networks have found their way into the research area of tropical cyclones. To the knowledge of the authors, Lee and Liu \cite{Lee00} firstly used satellite images of tropical cyclones for track prediction with the help of neural networks. But the images itself did not function as input data for the network. In a pre-processing step, information like the Dvorak number, the maximum wind speed, or the cyclone's position have been extracted from the images and fed to the network for a time-series prediction. The model has shown improvements of $30$ and $18~\%$, respectively, over the existing numerical model for forecasting tropical cyclone patterns and tracks. Kovordanyi and Roy \cite{Kovordanyi09} were the first who actually used satellite images as input data for a neural network. The network successfully detected the shape of a cyclone and predicted the future movement direction. Hong et al. \cite{Hong17} utilized multi-layer neural networks to predict the position of the cyclone's eye in a single high-resolution 3D remote sensing image. The network learns the coordinates of the eye from labeled images of past data and predicts them in test images. Kordmahalleh et al. \cite{Moradi16} have used a sparse recurrent neural network (RNN) to predict the trajectories of cyclones coming from the Atlantic Ocean or the Caribbean Sea, also known as hurricanes. They have used dynamic time warping (DTW), which lets the neural network learn information equally from each hurricane. Alemany et al. \cite{Alemany18} have also used a RNN to predict hurricane trajectories, but instead of assuming monotonic behavior of all hurricanes they focus on temporal and unique features of each cyclone. By learning extracted parameters, like the angle of travel or wind speed, their forecast results could achieve a better accuracy than the results of Moradi Kordmahalleh et al.. Zhang et al. \cite{Zhang18} have utilized a matrix neural network for predicting the trajectories of cyclones that occurred in the South Indian ocean. They state that matrix neural networks (MNN) suit better to the task of cyclone trajectory prediction than RNNs, because MNNs can preserve spatial information of cyclone tracks. Giffard-Roisin et al. \cite{Giffard-Roisin18} have used reanalysis data in combination with an artificial neural network. Focusing only on the typhoon center coordinates, an averaged error of $32.9~km$ has been achieved. Li et al. \cite{Li18} used a generative adversarial network (GAN) to predict future cloud motion for a very short time scale. Images of one typhoon function as training data and images of another typhoon are used for testing.       

The current study is a continuation of a previous work done by the authors (see Rüttgers et al. \cite{Ruettgers18}). In the previous work, a GAN has been applied for predicting tracks of typhoons. A GAN is a deep learning technique used to generate samples in forms of images, music or speech \cite{Goodfellow14}. In the previous work, chronologically ordered satellite images from the past have been used as input and images that show the typhoon one time step ahead have been generated as output. The GAN has been trained with satellite images from 66 typhoons and tested with images from 10 typhoons that did not belong to the training data. The time gap between images and therefore the prediction time was $6$ hours. Typhoon center coordinates could be predicted with an average error of $95.6~km$. Furthermore, the GAN was able to predict overall trends in cloud motion. However, the predicted images suffered from blurriness.

In the current study, the satellite images used in the previous work are combined with meteorological data from a reanalysis database. Reanalysis data are historical meteorological data that are reconstructed with the help of numerical simulations. The data used in the current work are provided by the European Centre for Medium-Range Weather Forecasts (ECMWF) \cite{ECMWF}. They contain information that are physically meaningful for the development of typhoons, for example the sea surface temperature (SST) or surface pressure (SP).     

The paper is organized as follows. The observational and meteorological datasets used as input data for the GAN are presented in section 2. The deep learning methodology is explained in section 3. This includes the architecture and the loss function of the GAN. In the fourth part the prediction results of the current study are compared to the results of the previous work done by the authors. This is followed by concluding remarks in the last part.

\section{Datasets}

\subsection{Observational data}

The satellite images have been provided by the Korean Meteorological Administration (KMA) \cite{KMA}. They contain those $76$ typhoons from $1993$ till $2017$ that hit or were about to hit the Korean peninsula. During the $25$ years of capture time, different satellites have been operating. Detailed information about types of satellites and pre-processing steps that are necessary before being used in the GAN are given in the authors' previous work (see Rüttgers et al. \cite{Ruettgers18}).

All input satellite images have a size of $250$$\times$$238$ pixels and three color channels (R(red)G(green)B(blue)). In total $1,628$ images are stored, with a time step size of $6$ hours between the images. There are two accuracy criteria for the typhoon track prediction, a quantitative and a qualitative one. For the quantitative criterion the difference between the predicted typhoon center coordinates and the real ones are taken into consideration. In the qualitative criterion the future shape of the clouds is estimated. 

\begin{figure}
  \centerline{\includegraphics[scale=0.45]{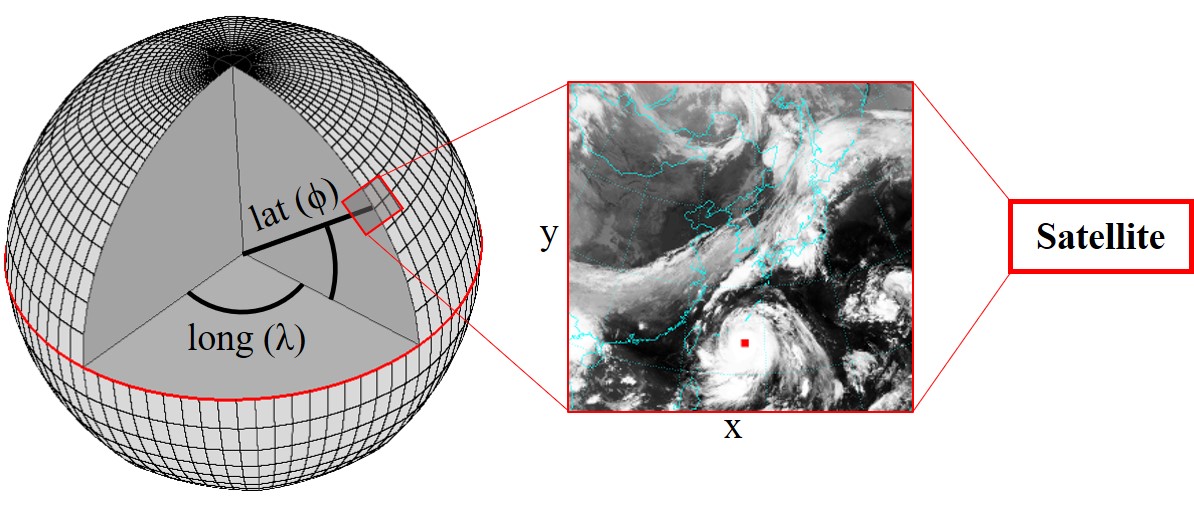}}
  \caption{Georeferencing of the typhoon center coordinates.}  
\label{fig:georeferencing}
\end{figure}

Every satellite image has been labeled with a red square at the typhoon center. The latitudinal ($\phi$) and longitudinal ($\lambda$) coordinates of the typhoon centers have been provided by the Japan Meteorological Agency (JMA) \cite{JMA}. In order to label each satellite image with its red square, the $\phi$ and $\lambda$ coordinates have to be transferred to (x,y)-pixel coordinates in the image. This process is known as georeferencing and is illustrated in figure~\ref{fig:georeferencing}.

The images are split into training and test data. The training data contain $1,389$ images of $66$ typhoons, the test data are $239$ images of $10$ typhoons. This study focuses on the development of each typhoon until it reaches the land. Typhoon centers and cloud motion after landfall are not considered. 

\subsection{Meteorological data}

In general, the ECMWF provides global forecasts, climate reanalyses and specific datasets, designed to meet different user requirements. The requirement for this work is to obtain data that help to learn the movement of typhoons. In order to fulfill this requirement, data of the public dataset ERA-interim are chosen \cite{ETA-interim}. ERA-interim is a global atmospheric reanalysis starting from 1979 that has continuously been updated until today. It uses a fixed version of a numerical weather prediction system (IFS - Cy31r2) to produce reanalyzed data. The available data have the following characteristics \cite{Dee11}:

\begin{itemize}
	\item Time step size: $6$ hours
	\item Spatial coverage: Global
	\item Lat/long grid resolution: $0.75~deg$ ($\approx~83 km$)
	\item Vertical levels: $60$ levels from the surface to $0.1~hPa$
\end{itemize}  

The downloaded data have time steps of $6$ hours and their dates fit to the satellite images described in the previous section. Although ERA-interim data cover the whole globe, for this work only the area around the Korean peninsula is selected. Raw data with a grid resolution of $0.75$ degrees are refined to $0.125$ degrees by applying linear interpolation, which leads to a resolution of nearly $13.8~km$. Georeferencing is done with the software Panoply and the map type Lambert Conformal Conic to match the view of the satellite images \cite{Panoply}. From reanalysis data, images are generated and cropped to the same size than the satellite images. The data contain the following physical quantities, which have direct impacts on the development of typhoon movements.

\textbf{Sea surface temperature (SST)}

Typhoons get their energy from water that evaporates at the sea surface, rises, and condensates at high altitudes. Through condensation latent heat gets released, that feeds typhoons with energy. Warm water at the sea surface ensures high evaporation rates, a higher water content in the air and more energy for typhoons. Thus, the sea surface temperature highly affects the development of such cyclones. At temperatures below $26.5^\circ$C, the buoyant force is not strong enough and sinking overcomes the upward motion, also known as convection. In general, a SST threshold of $26.5^\circ$C is necessary for strong convection, so that enough moist air can rise and tropical cyclones form \cite{Tory15}. A SST above $29.0^\circ$C, however, can have a negative effect on convection \cite{Lau97}. 

\begin{figure}
  \begin{center}
  \centerline{\includegraphics[scale=0.4]{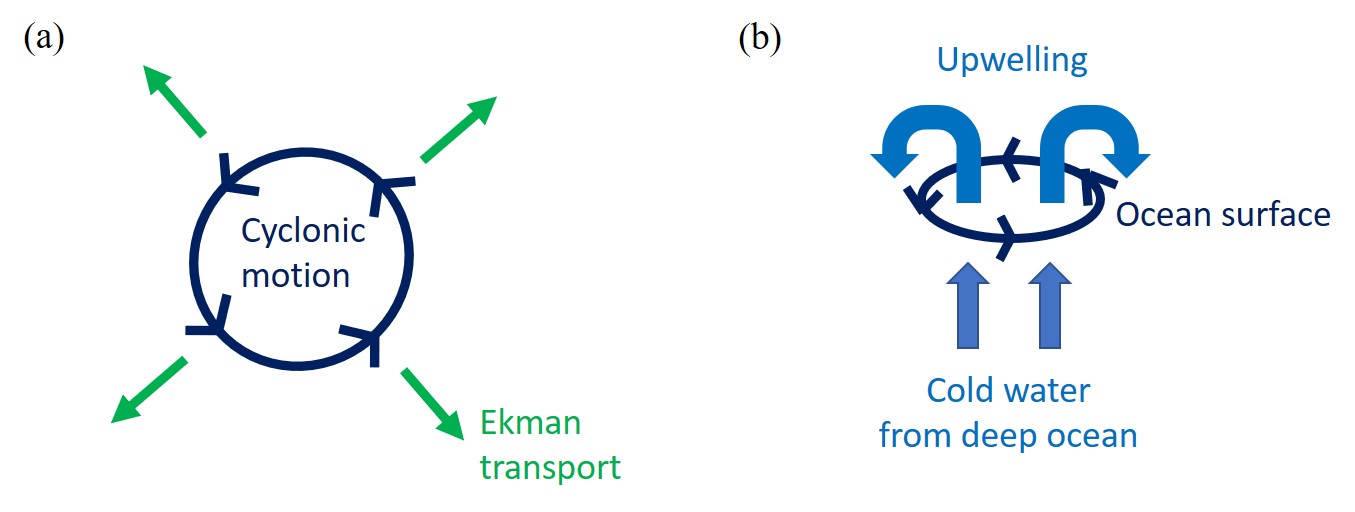}}
  \end{center}  
\caption{Ekman transport (a) and upwelling (b) shown on a simplified typhoon model.}  
\label{fig:cyclone}
\end{figure}

But this describes only the interaction between the SST and how typhoons develop in the atmosphere. There is also an ocean response on the SST. Typhoons spin in cyclonic (counter-clockwise) direction, as shown in figure~\ref{fig:cyclone}(a). Their spinning motion couples with the surface layer of the ocean. The scale of a typhoon is large enough to be affected by the Coriolis force of the spinning earth. In the northern hemisphere, this force causes a deflection of $90^\circ$ to the right. Thus, the surface layer also gets deflected, which is known as Ekman transport. The Ekman transport causes upwelling at the ocean surface and cold water gets sucked from the deep ocean, as illustrated in figure~\ref{fig:cyclone}(b). This means, wherever a typhoon moves, it leaves colder water behind.

\begin{figure}
  \begin{center}
  \centerline{\includegraphics[scale=0.34]{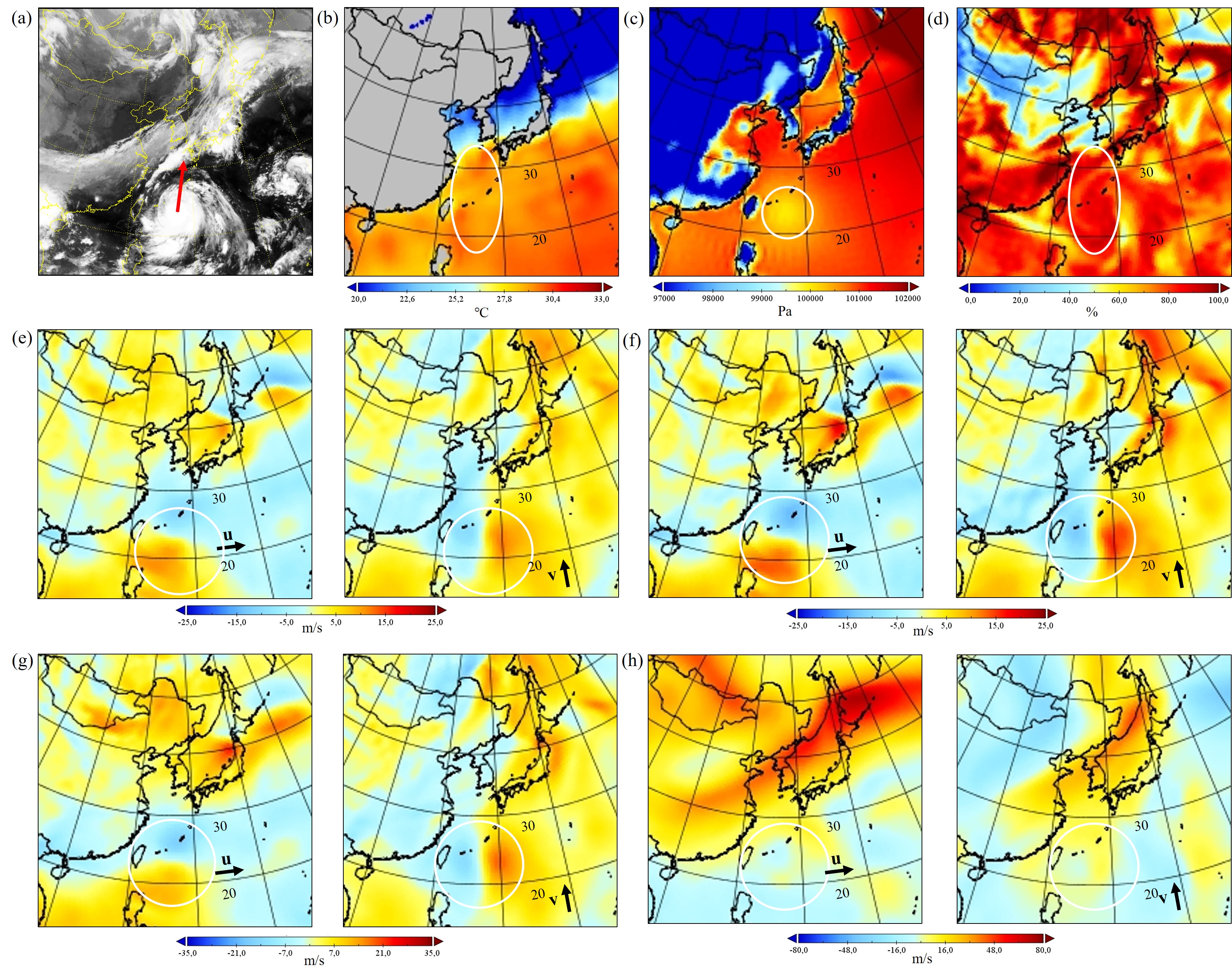}}
  \end{center}  
\caption{Satellite image (a), sea surface temperature (SST) (b), surface pressure (SP) (c) and relative humidity (RH) (d), Zonal [u] and meridional [v] surface velocity (e), velocity field at $950~mb$ pressure level (f), velocity field at $850~mb$ (g) and velocity field at $200~mb$ (h) of typhoon Maemi on September 10th in 2003 at 0:00 Coordinated Universal Time (UTC).}  
\label{fig:reanalysis}
\end{figure} 

As an example, figure~\ref{fig:reanalysis} shows the satellite image (a) and the SST reanalysis image (b) of typhoon Maemi on September 10th in 2003 at 0:00 Coordinated Universal Time (UTC). The red arrow in figure~\ref{fig:reanalysis}(a) indicates the moving direction of the typhoon. The example in figure~\ref{fig:reanalysis}(b) shows both, the SST-atmosphere relation and the SST-ocean response. Inside of the white ellipse near the Korean and Japanese coasts one can see the warm sea surface that strengthened Maemi when moving northwards from its current position. This was one reason for the disastrous strength of the typhoon near the coast \cite{Yun12}. On the other hand, due to the upwelling effect, the SST at the current position is cooler than the SST at its surroundings.

The SST in the input data for the GAN ranges from $15^\circ$C to $33^\circ$C. As an exception, instead of $6$ hour intervals only the daily average SST was available.

\textbf{Surface pressure (SP)}

The area surrounding the Korean peninsula lies within a so called Hadley cell. Warm, moist air rises at the equator and is transported from the equator poleward to about the $30^{th}$ latitude. Here, the now cool and dry air sinks and is responsible for a high surface pressure band, called the subtropical high. In figure~\ref{fig:reanalysis}(c) the pressure at the sea surface along this band reaches up to $102,000~Pa$. Typhoons, however, are known to produce local low pressure areas in their center, the eye of the typhoon. These areas are caused by rising air that transports evaporated water to high altitudes, which has been described previously in the text. Once humid air reaches those altitudes, water condensates and latent heat is released into the atmosphere. This heating causes air to expand and diverge at upper levels, which sucks air from lower levels and reduces pressure there. The white circle in figure~\ref{fig:reanalysis}(c) clearly highlights the low SP area in the center of typhoon Maemi. The input data for the GAN have a range from $97,000$ to $102,000~Pa$. Since the focus lies on the sea surface, pressure areas below $97,000~Pa$ at the mainland are not taken into consideration. 

\textbf{Relative humidity (RH)}

The size of a tropical cyclone highly depends on the environmental humidity. Moister environments are associated with larger storms, since more latent heat can be released from condensed water. A higher amount of condensed water combined with diverged air at upper levels causes more precipitation in outer rainbands, which leads to a lateral expansion of the tropical cyclone wind field \cite{Hill09}. Dry air, on the other hand, limits the strengthening of typhoons. In figure~\ref{fig:reanalysis}(d) relative humidity is illustrated for the case of typhoon Maemi at a height in terms of pressure level of $850~mb$. RH is the ratio of the partial pressure of water vapor to the equilibrium pressure of a vapor-air mix. A percentage value shows the ratio of the current water vapor load to the maximum load of water vapor. A relative humidity of $100~\%$, for example, means the air is carrying the maximum load of water vapor. The white ellipse in figure~\ref{fig:reanalysis}(d) covers the area of impact of typhoon Maemi, as well as the area of its future movement. The cyclone is continuously fed with moist air, which was another reason for its enormous strength.

\textbf{Surface velocity field}

To analyze the near surface velocity, the zonal (u) and meridional (v) velocity components at $10$ meter height have been extracted (see figure~\ref{fig:reanalysis}(e)). Surface velocities are used to measure the intensity of a tropical cyclone. According to its intensity a typhoon gets classified from class 1, a tropical depression, to class 6, a violent typhoon. But the surface velocity is also important for the movement of typhoons, because it reflects information about the surface topology, for example the impact of islands on typhoon tracks. The GAN should learn such geographical intricacy. In figure~\ref{fig:reanalysis}(e) surface velocities near the typhoon center are found inside of the white circles. One can imagine how the combined zonal and meridional components create strong local vorticities and give the typhoon its spinning motion.

\textbf{Velocity field at $950~mb$ pressure level}

Just learning from the surface velocity profile is not sufficient to describe the motion of a typhoon. Tropical cyclones are characterized by different maximum speeds at different lateral positions and different heights. The eyewall of a cyclone is a ring that surrounds the eye. It is the most devastating area. Measurements of GPS dropwindsondes have revealed that hurricanes have their strongest mean eyewall wind speeds at a height of around $500~m$ \cite{Franklin02}. These strong winds are taken into consideration for the learning phase of the GAN. Among the available pressure levels from the ERA interim dataset, a level of $950~mb$ comes closest to an altitude of $500~m$. Figure~\ref{fig:reanalysis}(f) gives insights about the velocity field at $950~mb$ for Maemi. The velocity distribution near the typhoon is similar to the surface velocity distribution, except for higher velocity magnitudes at $950~mb$.

\textbf{Vertical wind shear}

\begin{figure}
  \begin{center}
  \centerline{\includegraphics[scale=0.5]{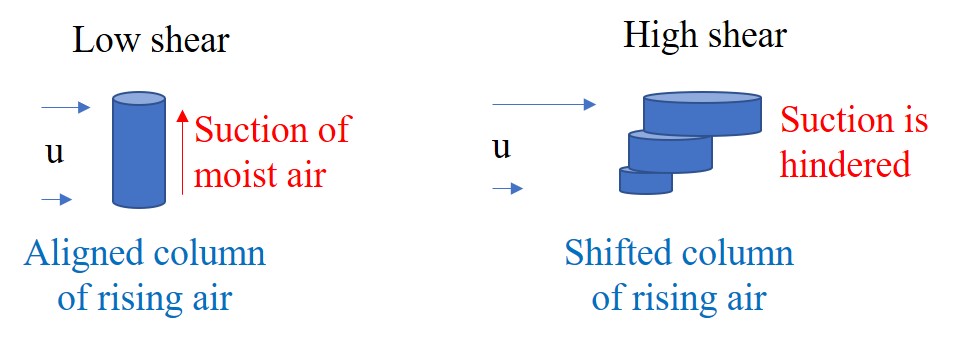}}
  \end{center}  
\caption{Vertical wind shear explained on a simplified model.}  
\label{fig:shear}
\end{figure}

In strong typhoons moist air rises through a column from the surface to high altitudes. Low vertical wind shear strengthens this column and the suction effect inside, since lower and upper parts of the column are aligned. A storm's released latent heat is then concentrated on a relatively small area. High vertical wind shear, on the other hand, makes it difficult for the column to keep all parts aligned and create a strong suction effect. This phenomenon is illustrated in figure~\ref{fig:shear}. Together with data of the previously introduced surface velocity field and velocity at a pressure level of $950~mb$, data of the velocity fields at $850~mb$ and $200~mb$ pressure level are generated to consider vertical wind shear in the learning process. Figures~\ref{fig:reanalysis}(g) and (h) show the velocity field of Typhoon Maemi at $850~mb$ and $200~mb$ pressure levels. Those two altitudes are commonly used when analyzing environmental wind shear and its effect on tropical cyclones \cite{Chen06}. The velocity range at $200~mb$ pressure level is much wider, because large scale winds with high velocities have to be considered, for example the jet stream in figure~\ref{fig:reanalysis}(g) above the Korean peninsula.     

All presented physical quantities are not just affecting the development of typhoons directly, they are also highly interdependent. Surface winds, for example, can be influenced by the SST. Above an area with cold SST the air flow is more stable than above an area with warm SST. Stable airflow induces weak winds, low water air mixing and low evaporation, which weakens typhoons. Unstable airflow, on the contrary, favours evaporation and thus the development of tropical cyclones. Such interdependencies are learned by the GAN.

In total, $15,279$ training data and $2,629$ test data are gained from the reanalysis database. They are generated as grey scale images.

\section{Deep learning methodology}

\begin{figure}
  \begin{center}
  \centerline{\includegraphics[scale=0.5]{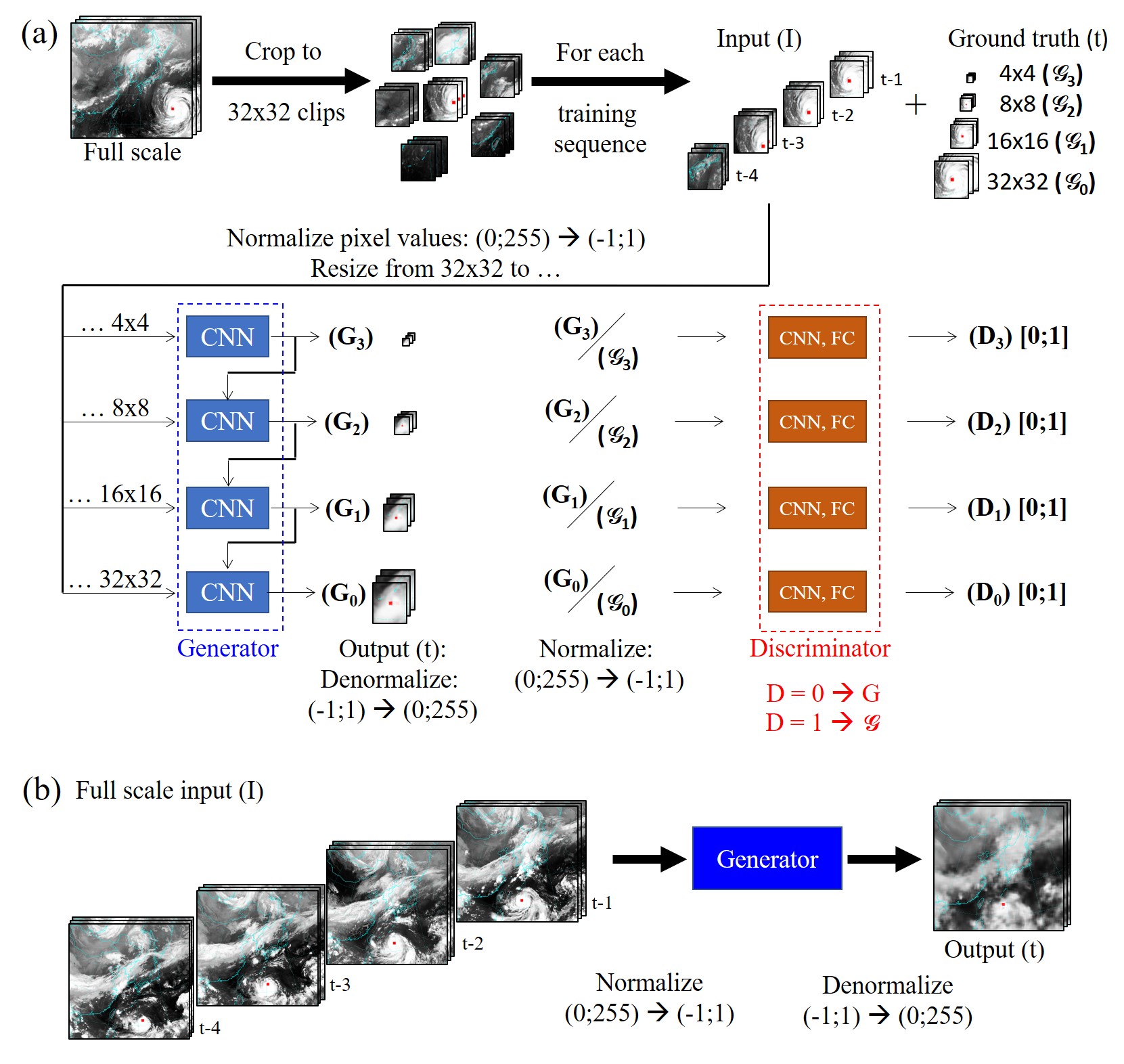}}
  \end{center}  
\caption{Training (a) and testing (b) of the GAN.}  
\label{fig:gan_1}
\end{figure}

The deep learning methodology is different for training and testing steps. Figure~\ref{fig:gan_1} (a) shows the concept of training steps. For the simplicity, a case is shown where only satellite images function as input. As mentioned previously in this work, each full scale satellite image consists of three color channels (RGB). They are represented throughout figure~\ref{fig:gan_1} by three overlapped layers. Before using the training data as input images to the GAN they get cropped to a total number of $5,000,000$ clips with a pixel size of $32$$\times$$32$. One clip contains a set of $m$ consecutively cropped satellite images from the past (input, $I$) and one ground truth image. On the one hand, such a high number of training images reduces the risk of overfitting. On the other hand, using cropped parts reduces the memory used during one training step. Additionally, focusing on small parts of an image allows to learn the details of the image better. As an example, in figure~\ref{fig:gan_1} (a) four input images are chosen ($m=4$). 

The cropped training clips are trained using variations of deep learning networks in a GAN based video modeling architecture \cite{Mathieu15}. This architecture contains a pure convolutional neural network, the so called generator, and a network that combines convolutional layers with fully connected layers, the so called discriminator. The generator network takes a clip, normalizes its pixel values from ($0;255$) to ($-1;1$), predicts a $32$$\times$$32$ part of a satellite image at future occasion (output, $G_{0}$) and denormalizes the predicted pixels from ($-1;1$) to ($0;255$). GANs are multi-scale networks, thus the generator network generates images on different scales ($G_{0}$,$G_{1}$,$G_{2}$,$G_{3}$). The generated images with pixel sizes of $4$$\times$$4$, $8$$\times$$8$ and $16$$\times$$16$ function as additional input for the next bigger scale, as pointed out in figure~\ref{fig:gan_1} (a). This helps to keep long-range dependencies and preserves high resolution information throughout the convolutional learning. 

The discriminator network, on the other hand, tries to classify between the generated and ground truth images of each scale. The output layer provides a binary output ($D$), where $0$ stands for a generated image and $1$ means ground truth. Generator and discriminator are trained in alternated steps with a number of $mbs$ (mini batch size) clips for the generator step and $mbs$ new clips for the discriminator steps. Optimizing both networks in the same loop with the same clips would lead to overfitting \cite{Goodfellow14}.  

Testing steps are done with full scale test data, containing images with $250$$\times$$238$ pixels. They are not cropped. As visualized in figure~\ref{fig:gan_1} (b), input images are taken and normalized by the generator network which generates a full scale image at a future occasion and denormalizes it. Again, as an example four input images are chosen ($m=4$).  

When combining satellite images with reanalysis data, instead of only $3$ RGB channels there will be $14$ input channels. The additional $11$ channels come from the physical quantities that have been introduced earlier in this study.   

When the discriminator tries to classify, the generator tries to fool the discriminator and makes it hard to classify. The fooling can be understood by looking at the loss functions. The generator model is trained to minimize a combination of loss functions as
\begin{eqnarray}
{L}_{generator} = 1/4 \sum_{k=0}^{3}{\lambda_{l2} {L}_{2}^{k} + \lambda_{gdl} {L}_{gdl}^{k} + \lambda_{adv} {L}_{adv}^{G,k}},
\end{eqnarray}
where $\lambda_{l2} = 1$, $\lambda_{gdl} = 1$, and $\lambda_{adv} = 0.05$.

Let $G_{k}(I)$ be the generated image and $\mathcal{G}_{k}(I)$ be the ground truth image resized by $\frac{1}{2^{k}}$. The ${L}_{2}^{k}$ loss function evaluates the explicit difference between the predicted and provided images on each scale as
\begin{eqnarray}
{L}_{2}^{k} = ||G_{k}({I}) - \mathcal{G}_{k}({I})||_{2}^{2}.
\end{eqnarray}

The $\mathcal{L}_{gdl}^{k}$ loss function compares the difference between gradients of the predicted and provided images as
\begin{eqnarray}
{L}_{gdl}^{k} =  \sum_{i=0}^{n_x-2} \sum_{j=0}^{n_y-2} \Big\{ \bigg| |\mathcal{G}_{k}({I})_{(i+1,j+1)}-\mathcal{G}_{k}({I})_{(i,j+1)}| -|G_{k}({I})_{(i+1,j+1)} - G_{k}({I})_{(i,j+1)}| \bigg| \\ \nonumber
+\bigg||\mathcal{G}_{k}({I})_{(i+1,j+1)}-\mathcal{G}_{k}({I})_{(i+1,j)}|  -|G_{k}({I})_{(i+1,j+1)} - G_{k}({I})_{(i+1,j)}| \bigg| \Big\},
\end{eqnarray}
where $n_x$ and $n_y$ are numbers of pixel in the width and the height of an image, and ($i,j$) is a pixel coordinate. This helps to sharpen the prediction by directly penalizing the difference of gradients between neighboured pixels.

The $\mathcal{L}_{adv}^{G,k}$ loss function is applied to support the generator model deluding the discriminator network by generating images which are indistinguishable from the provided ground truth as
\begin{equation}
{L}_{adv}^{G,k} = L_{bce}(D_{k}(G_{k}({I})),1),
\label{eqn: ladvg}
\end{equation}
where $L_{bce}$ is the binary cross entropy loss function defined as
\begin{equation}
L_{bce}(a,b) = -b \log(a) - (1-b) \log(1-a),
\label{eqn: lbce}
\end{equation}
for scalars $a$ and $b$ between $0$ and $1$. The loss gets smaller when the discriminator identifies the generated images as a ground truth, which is a wrong classification.

The discriminator model is trained to minimize a loss function as
\begin{equation}
{L}_{discriminator} = \frac{1}{4}\sum_{k=0}^{3} \left[ L_{bce}(D_{k}(\mathcal{G}_{k}({I})),1) + L_{bce}(D_{k}(G_{k}({I})),0)\right].
\label{eqn: ladvd}
\end{equation}
This supports the network not to be deluded by the generator model by extracting important features of typhoons in an unsupervised manner. Here, the loss gets smaller when the discriminator classifies correctly.

The configuration of the deep learning network is summarized in table~\ref{tab:network} (see \cite{Mathieu15} for detailed algorithms).
As shown in figure~\ref{fig:gan_1}, the generator model is comprised of four fully convolutional networks with multi-scale architectures. As mentioned before, the number of channels (ch) can be $3$ in case of satellite images only and $14$ in case of a combination of satellite images and images from reanalysis data. The discriminator model of the deep learning network is comprised of a sequence of multi-scale networks ($D_{0},D_{1},D_{2},D_{3}$) with convolutional layers, a pooling layer and fully connected layers. 

\begin{table}[h]
        \begin{center}
                \def~{\hphantom{0}}
                \begin{tabular}{c|l|l}
                    \hline \hline
                    \multicolumn{3}{c}{Generator model}\\ \hline \hline
                        \multicolumn{3}{c}{Convolution layer}\\ \hline
                        Network & Feature map sizes & Kernel sizes\\ \hline
                        $G_{3}$ & ch$\times$ m ~128 ~256 ~128 ~ch & $3\times3$, $3\times3$, $3\times3$, $3\times3$ \\
                        $G_{2}$ &ch$\times$ (m+1) ~128 ~256 ~128 ~ch & $5\times5$, $3\times3$, $3\times3$, $5\times5$ \\
                        $G_{1}$ &ch$\times$ (m+1) ~128 ~256 ~512 ~256 ~128 ~ch & $5\times5$, $3\times3$, $3\times3$, $3\times3$, $3\times3$, $5\times5$  \\
                        $G_{0}$ & ch$\times$ (m+1) ~128 ~256 ~512 ~256 ~128 ~ch & $7\times7$, $5\times5$, $5\times5$, $5\times5$, $5\times5$, $7\times7$ \\  \hline\hline
                        \multicolumn{3}{c}{Discriminator model}\\ \hline \hline
                        \multicolumn{3}{c}{Convolution layer}\\ \hline
                        Network & Feature map sizes & Kernel sizes\\ \hline
                        $D_{3}$ & ch 64 & $3\times3$ \\
                        $D_{2}$ &ch 64 128 128 & $3\times3$, $3\times3$, $3\times3$ \\
                        $D_{1}$ &ch 128 256 256 & $5\times5$, $5\times5$, $5\times5$  \\
                        $D_{0}$ &ch 128 256 512 128 & $7\times7$, $7\times7$, $5\times5$, $5\times5$ \\  \hline
                        \multicolumn{3}{c}{$2\times2$ max pooling}\\ \hline
                        \multicolumn{3}{c}{Fully connected layer}\\ \hline
                        Network & \multicolumn{2}{l}{Neuron numbers}\\ \hline
                        $D_{3}$ & \multicolumn{2}{l}{512 256 1} \\
                        $D_{2}$ & \multicolumn{2}{l}{1024 512 1} \\
                        $D_{1}$ & \multicolumn{2}{l}{1024 512 1} \\ 
                        $D_{0}$ & \multicolumn{2}{l}{1024 512 1} \\ \hline \hline
                \end{tabular}
                \caption{Configuration of the deep learning networks with the number of input channels ($ch$) and the number of input images ($m$).}
                \label{tab:network}
        \end{center}
\end{table}

 In the present study, ReLU activation functions are used for both, the generator and discriminator network. The output layer of the generator network cannot use ReLU, since its values need to be between $-1$ and $1$. Therefore, the tanh activation function is applied in this layer.

\section{Results and Discussion}

The GAN has been trained and tested on a NVIDIA Tesla K40c GPU. During training a mini batch size of $mbs=8$ and learning rates of $0.00004$ and $0.02$ have been chosen for the generator and discriminator network. These values have proven themselves to be valuable in the work of Mathieu et al.\cite{Mathieu15}, as well as in the study of Lee and You \cite{Lee18}.

The ten typhoons of the test dataset are: Faye (Juli $1995$), Violet (September $1996$), Oliwa (September $1997$), Saomai (September $2000$), Rammasun (June/July $2002$), Maemi (September $2003$), Usagi (July/August $2007$), Muifa (July/August $2011$), Neoguri (June/July $2014$), Malakas (September $2016$). The testing for each typhoon is done in sequences. One sequence contains the chronologically ordered input images, the ground truth image and the generated image. A sequence is named with the date and time (universal time coordinated - UTC) of the corresponding ground truth image, for example: "1993072718" stands for $27^{th}$ of July in $1993$ at $6~pm$ (UTC).

\subsection{Predicting typhoon center coordinates}

The prediction accuracy is estimated with the help of an absolute error $E$ and a relative error $E_{rel}$. $E$ is calculated in kilometers ($km$) by applying the haversine formula \cite{Mahmoud16}, with the earth radius $R$ taken at the location of the real coordinates of the ground truth of each sequence:

\begin{equation}
	E = 2R\arcsin{\sqrt{\sin({\frac{\phi_{pred}-\phi_{real}}{2}})^2+\cos{\phi_{real}}\cos{\phi_{pred}}\sin({\frac{\lambda_{pred}-\lambda_{real}}{2}})^2}}.
\end{equation}

$E_{rel}$ considers the ratio between $E$ and the distance that a typhoon has traveled over the last six hours, where $t$ stands for the time of a certain sequence: 

\begin{equation}
	\resizebox{0.95\hsize}{!}{$E_{rel}(t) = \frac{E}{2R\arcsin{\sqrt{\sin({\frac{\phi_{real}(t)-\phi_{real}(t-6h)}{2}})^2+\cos{\phi_{real}(t)}\cos{\phi_{real}(t-6h)}\sin({\frac{\lambda_{real}(t)-\lambda_{real}(t-6h)}{2}})^2}}}.$}
\end{equation}

\begin{figure}
  \centerline{\includegraphics[scale=0.64]{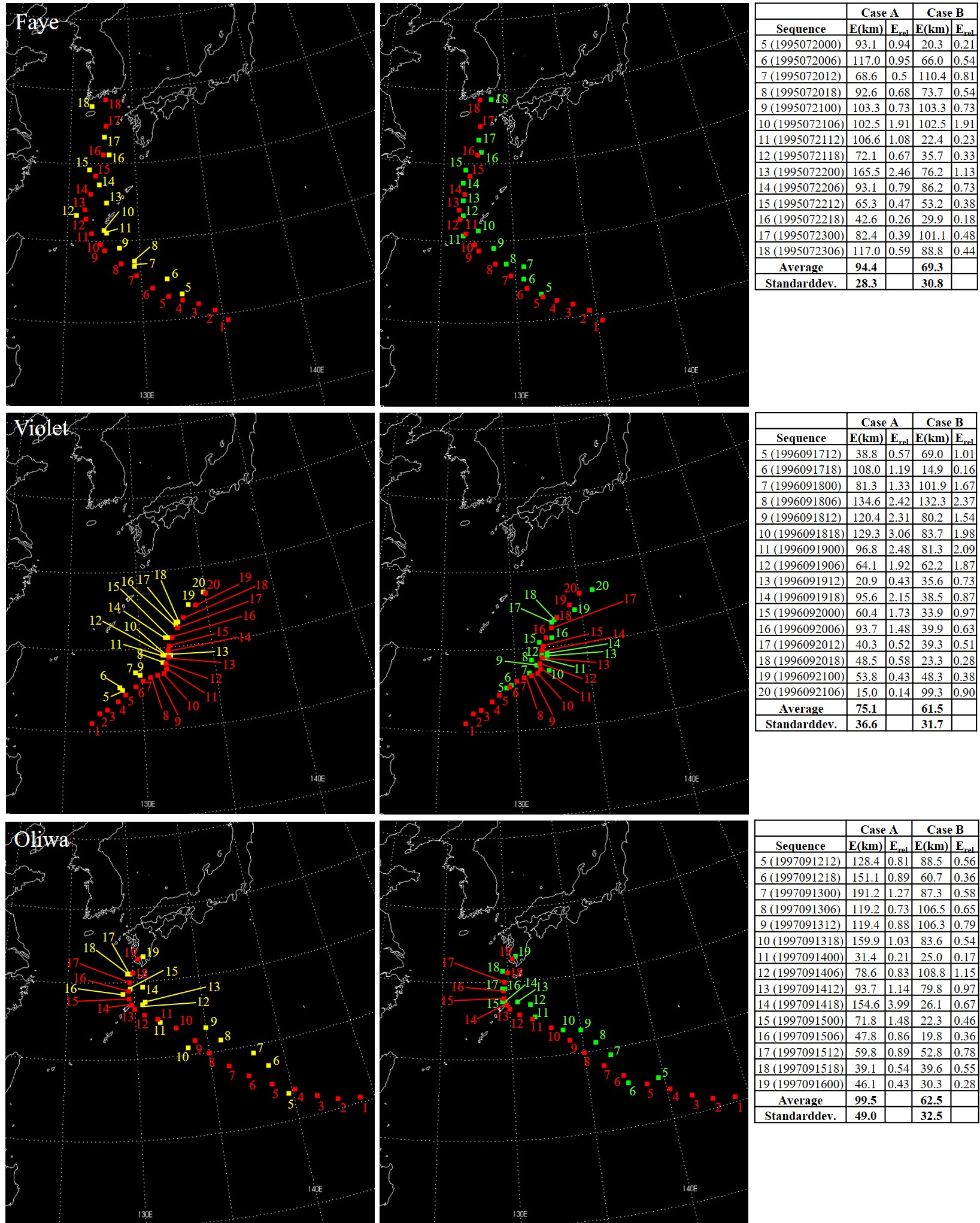}}
  \caption{Trajectories, absolute errors and relative errors for typhoons Faye, Violet and Oliwa. Red: Real typhoon center coordinates - Yellow: Predictions with $4$ input images, using only satellite images as input (Case A) - Green: Predictions with $4$ input images, using satellite images and reanalysis data as input (Case B).}  
\label{fig:results_4_input}
\end{figure}

\begin{figure}
  \centerline{\includegraphics[scale=0.64]{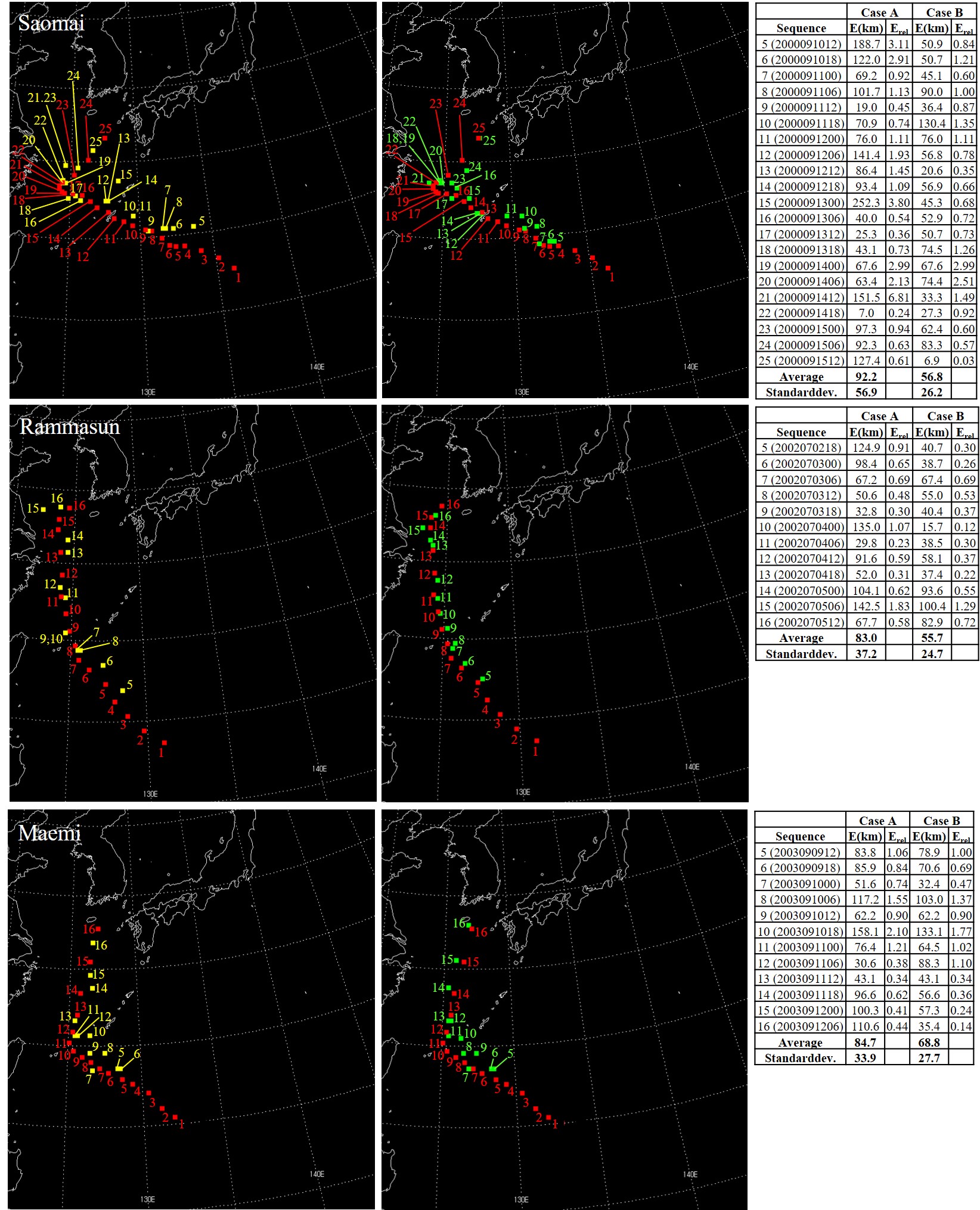}}
  \caption{Trajectories, absolute errors and relative errors for typhoons Saomai, Rammasun and Maemi. Red: Real typhoon center coordinates - Yellow: Predictions with $4$ input images, using only satellite images as input (Case A) - Green: Predictions with $4$ input images, using satellite images and reanalysis data as input (Case B).}  
\label{fig:results_4_input_1}
\end{figure}

\begin{figure}
  \centerline{\includegraphics[scale=0.64]{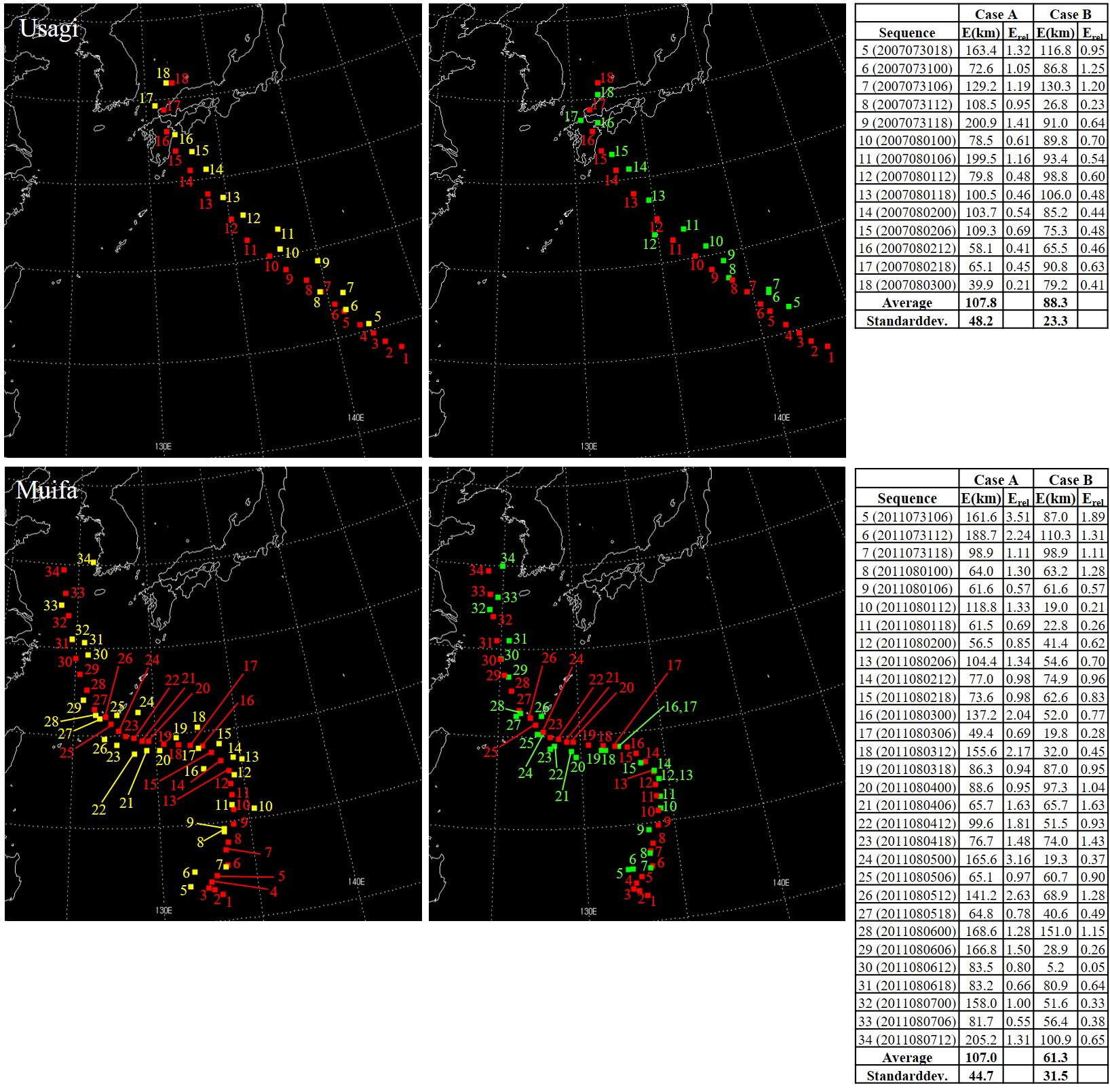}}
  \caption{Trajectories, absolute errors and relative errors for typhoons Usagi and Muifa. Red: Real typhoon center coordinates - Yellow: Predictions with $4$ input images, using only satellite images as input (Case A) - Green: Predictions with $4$ input images, using satellite images and reanalysis data as input (Case B).}  
\label{fig:results_4_input_2}
\end{figure}

\begin{figure}
  \centerline{\includegraphics[scale=0.64]{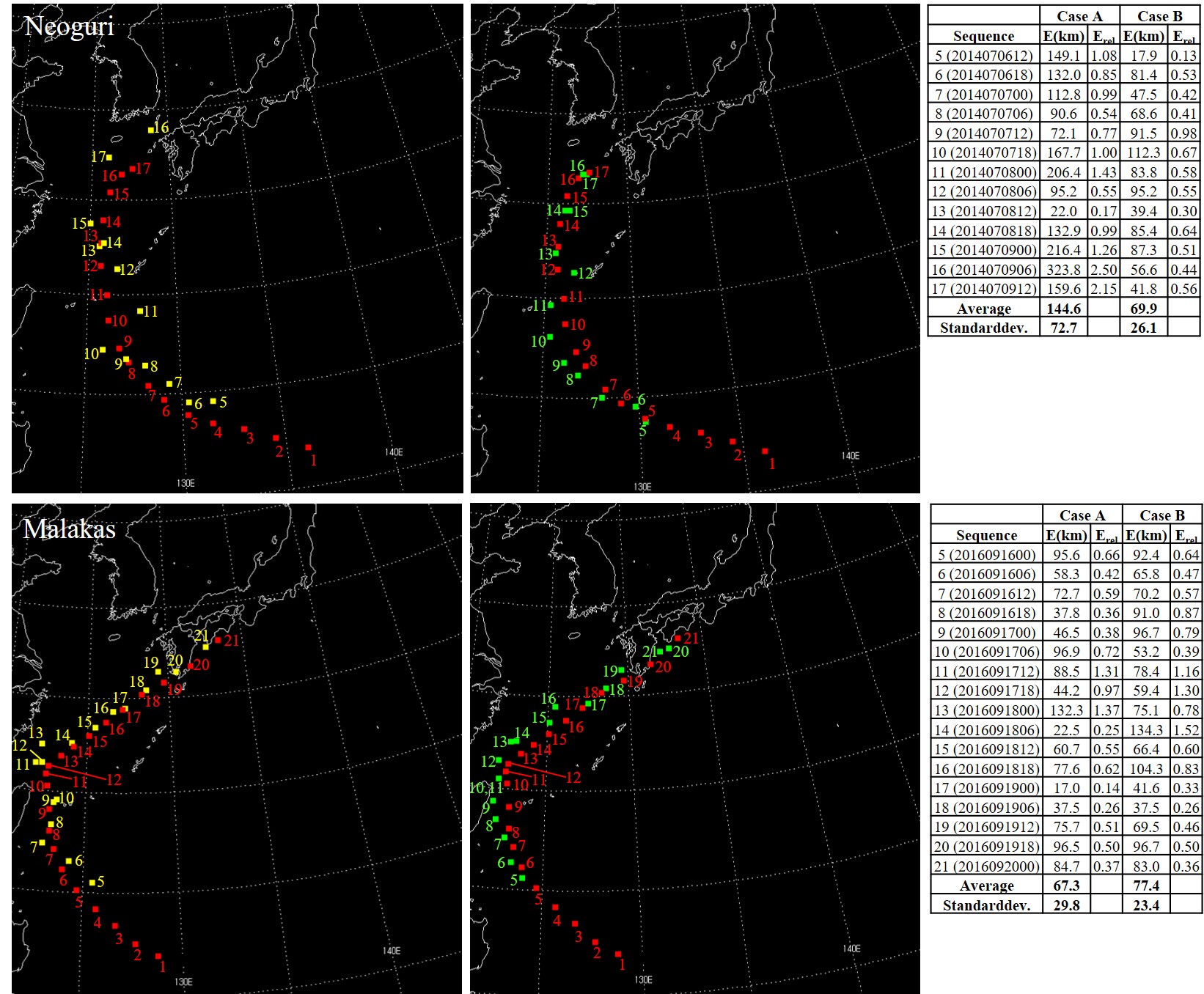}}
  \caption{Trajectories, absolute errors and relative errors for typhoons Neoguri and Malakas. Red: Real typhoon center coordinates - Yellow: Predictions with $4$ input images, using only satellite images as input (Case A) - Green: Predictions with $4$ input images, using satellite images and reanalysis data as input (Case B).}  
\label{fig:results_4_input_3}
\end{figure}

Figures~\ref{fig:results_4_input},~\ref{fig:results_4_input_1},~\ref{fig:results_4_input_2} and ~\ref{fig:results_4_input_3} present results of two different types on input data. The predictions in the left part of the figures are from the previous paper and have been made by using only satellite images as input data. They refer to case A and are marked with yellow color. The forecasts in the right part, on the other hand, are based on the combination of satellite images and reanalysis data. They refer to case B throughout the text and are colored green in the figures. The results of both cases are contrasted to the real typhoon center coordinates, represented by red squares. A parameter study in the previous work has shown that using four input images ($m=4$) delivers the best results \cite{Ruettgers18}. Thus, in both cases four input images have been used. 

Figure~\ref{fig:results_4_input} includes typhoons Faye, Violet and Oliwa. Faye reached the island Jeju on July $23^{rd}$ in $1995$. For case A, large errors are found at the initial sequences in the open water and at sequences $10$, $11$ and $13$, when the cyclone turns northward. In case B, the initial sequences and the direction change to the north are predicted much more accurately. The averaged absolute error ($E_{avg}$) is reduced by $26.6~\%$ from $94.4~km$ to $69.3~km$. Typhoon Violet was expected to become a serious threat to the Korean peninsula in its initial stage, but it changed its course towards Japan in the middle of September $1996$. Similar to Faye, when Violet changes its course at sequences $8-11$, the predictions become less accurate. Again, case B handles this situation better than case A. When the typhoon approaches the Japanese mainland the GAN predicts the typhoon center accurately in both cases. Cyclone Oliwa caused losses at off shore South Korea in September $1997$. For case A, the prediction quality is low when moving above the open water at sequences $5-10$ and when turning northwards at sequences $13-15$. Case B, on the contrary, shows improved results in both ranges of sequences, which leads to a reduction in averaged error by $37.2~\%$. Good results are achieved in both cases at the final sequences. 

Figure~\ref{fig:results_4_input_1} contains cyclones Saomai, Rammasun and Maemi. The track of typhoon Saomai shows two sudden course changes. At sequence $15$ it turned towards the Chinese mainland and was supposed to hit the area around Shanghai. Then, it suddenly changed its course again at sequence 21, moving towards South Korea and Japan. Both course changes cause problems to the prediction accuracy in case A. Case B, however, masters these sudden changes of direction. Furthermore, in case A the predicted centers show increased gaps above the open water at sequences $5$ and $6$, whereas these gaps do not occur in case B. The predictions based on satellite images and reanalysis data also allow a strong finish in the last sequence before landfall, resulting in a $39.4~\%$ decrease in $E_{avg}$. Typhoon Rammasun did not have significant course changes in summer $2002$. Only a minor course correction at sequence $10$ results in a slightly increased error in case A. The highest error in both cases is found before landfall at sequence $15$. The predictions for Maemi, one of the strongest and most dangerous typhoons that ever hit the Korean peninsula, again show a weakness when getting deflected northwards at sequences $8-11$. This is observed for both cases, though the predictions in case B are more accurate.

Figures~\ref{fig:results_4_input_2} and~\ref{fig:results_4_input_1} show the results for typhoons Usagi, Muifa, Neoguri and Malakas. No issues with changing course, but problems above the open water are observed for typhoon Usagi, especially at sequences $5$, $7$, $9$ and $11$. Once the cyclone approaches Japan and South Korea, the errors become more and more accurate. As seen before, predictions based on satellite images in combination with reanalysis data learn and predict the typhoon movement better than the forecasts made based on satellite images only. In August $2011$ the dragon shaped typhoon Muifa found its way to the Korean mainland. For case A, it is eye-catching how the GAN struggles initially at sequences $5-7$, as well as with sudden course changes at sequences $16-18$ or $24-26$. The highest absolute error is found at sequence $34$ right before the typhoon hits the Korean peninsula. Again, the overall performance of case B is much better with a reduction in $E_{avg}$ of $42.7~\%$. For case A, the predicted centers of typhoon Neoguri reveal very high errors before landfall at sequences $14-17$. These high inaccuracies are eliminated in case B. Here, adding reanalysis data leads to the highest reduction in average absolute error ($-51.8~\%$). The results for the youngest typhoon among the ten test cases, Malakas, show an exception. The average error in case B is slightly higher than in case A. However, the trajectory in case B looks more natural.

Except for typhoon Malakas, using reanalysis data in combination with satellite images improves the prediction performances for every test typhoon. The total averaged absolute error is $95.6~km$ for case A, compared to $67.2~km$ for case B, a reduction of $29.7~\%$. Furthermore, considering reanalysis data reduces the standard deviation of the error for all test typhoons, except for Faye. Thus, predictions do not only become more accurate, but also more stable. Trajectories are predicted more naturally.

Table~\ref{tab:frequency_distribution} provides the frequency distribution of $E$ for all sequences in both cases, A and B. In case A, $74.5~\%$ of the cases have an error of less than or equal to $120~km$. A combined $42.4~\%$ have an accuracy of less than or equal to $80~km$. The remaining errors are found above $120~km$. In case B, only five sequences remain in this class. A combined $65.5~\%$ have an accuracy of less than or equal to $80~km$.

\begin{table}[ht]
\centering
\begin{tabular}{|l|l|l|l|l|}
\hline
Error (km) & Case A & Percentage (\%) & Case B & Percentage (\%) \\
\hline
0-40 & 18 & 10.9 & 43 & 26.1 \\
\hline
41-80 & 52 & 31.5 & 65 & 39.4 \\
\hline
81-120 & 53 & 32.1 & 52 & 31.5 \\
\hline
$\geq$ 121 & 42 & 25.5 & 5 & 3.0 \\
\hline
\end{tabular}
\caption{Frequency distribution for the prediction of all sequences with $4$ input images, using only satellite images as input (Case A) and $4$ input images, using satellite images and reanalysis data as input (Case B).}
\label{tab:frequency_distribution}
\end{table}

In case A, most of the sequences with high errors deal with three major challenges. Firstly, it can be difficult to predict the motion above open water, when the typhoons have many possible paths to take and their path is not limited by land. Secondly, sudden changes in direction can cause increased relative errors. Finally, sometimes the relative error in sequences right before landfall is difficult to predict. The results of case B, however, have clearly shown that combining satellite images and reanalysis data as input for the GAN improves the prediction accuracy and helps to tackle these three challenges.

\subsection{Predicting cloud motion}

\begin{figure}
  \centerline{\includegraphics[scale=0.85]{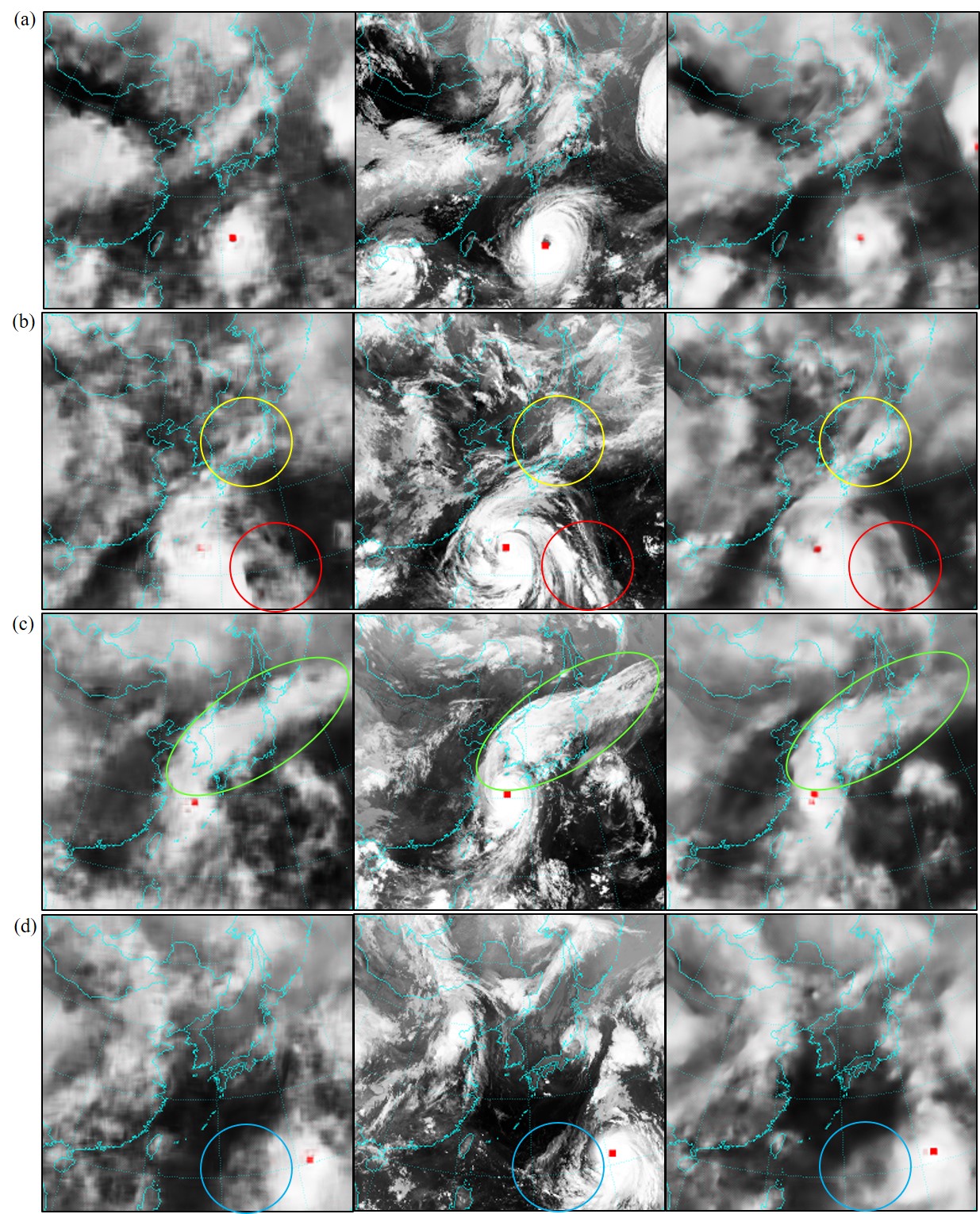}}
  \caption{Randomly chosen generated images for case A (left) and case B (right), compared to their corresponding ground truth image: (a) Sequence $11$ of typhoon Violet, (b) sequence $6$ of typhoon Rammasun, (c) sequence $15$ of typhoon Maemi and (d) sequence $5$ of typhoon Usagi.}  
\label{fig:cloud_motion_1}
\end{figure}

Cloud structures in satellite images are not represented in physical units. Thus, rather than measuring the accuracy of the cloud motion prediction quantitatively, this study focuses on predicting trends in cloud motions qualitatively. Figure~\ref{fig:cloud_motion_1} shows results from randomly picked sequences of case A on the left and case B on the right. They are compared to their corresponding ground truth image in the center. The figure contains images of sequence $11$ of typhoon Violet (a), sequence $6$ of typhoon Rammasun (b), sequence $15$ of typhoon Maemi and sequence $5$ of typhoon Usagi (d).

\begin{figure}
  \centerline{\includegraphics[scale=0.5]{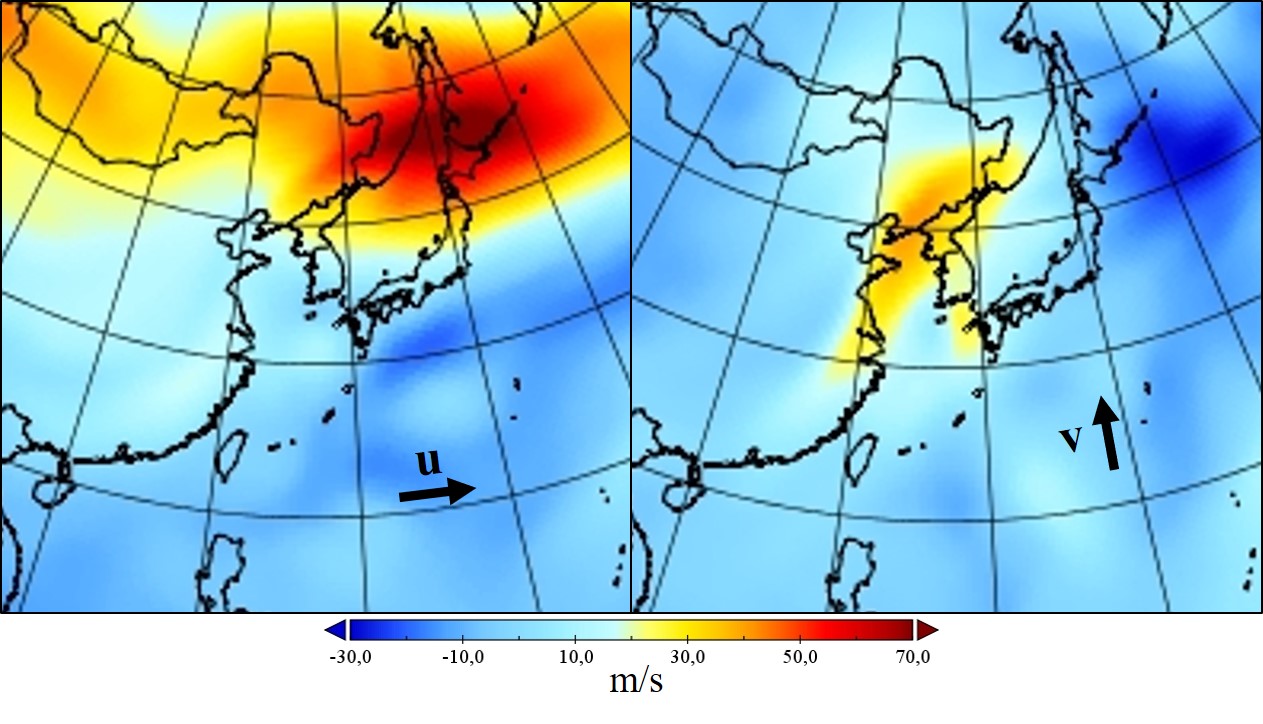}}
  \caption{Zonal [u] and meridional [v] velocity at $200~mb$ pressure level of typhoon Maemi on September $12^{th}$ in $2003$ at 6:00 Coordinated Universal Time (UTC).}  
\label{fig:cloud_motion_2}
\end{figure}

In general, the generated images do not have the same sharpness than the ground truth images, instead they are blurry. Blurriness is a well known challenge in video prediction tasks. The $\mathcal{L}_{gdl}$ loss function has been introduced to reduce blurriness \cite{Mathieu15}, but it could not totally be avoided in the present study. However, although the generated image suffers from a certain degree of blurriness, main differences between predictions of case A and B are clearly visible. The generated images of case A seem to be static and do not represent any dynamics. The generated images of case B, on the other hand, are much smoother. In sequence $11$ of typhoon Violet, for example, the image of case A cannot reproduce the spinning motion of the typhoon (see figure~\ref{fig:cloud_motion_1} (a)). The image of case B, on the contrary, does not only illustrate the typhoon more realistically, but also the cloud patterns in the remaining image. The GAN also recognizes the center of another cyclone that is moving at the right edge of the image. In sequence $6$ of typhoon Rammasun, the prediction of case B again resolves the spinning motion of the typhoon much better than the prediction of case A (see figure~\ref{fig:cloud_motion_1} (b)). Details, like the cloud structure east (red circles) or north (yello circles) of the typhoon center, are forecast much more reliably. In sequence $15$ of typhoon Maemi the cyclone has almost reached the Korean peninsula. Parts of the cloud structure start to dissipate, but a significant part follows a strong westerly wind at a high altitude, called jet stream. The jet stream with high zonal velocities of about $70~m/s$ is shown more clearly in figure~\ref{fig:cloud_motion_2}. In figure~\ref{fig:cloud_motion_1} (c) it can be noticed how the result of case B reproduces the suction effect of the jet stream in a smoother way than the predicted image of case A (green ellipses). Finally, in figure~\ref{fig:cloud_motion_1} (d) the image on the right highlights much better how surrounding clouds move into the eyewall than the image on the left (blue circles). In total, it is shown how blurriness can be reduced by adding physically meaningful information to the input data.      

\section{Conclusion}

The application of a data-driven deep learning method for typhoon track prediction in forms of typhoon center coordinates and cloud structures has been explored. Initially, only satellite images have been used to generate images that show future motion of typhoons. $4$ input images turned out to be the best choice and an overall typhoon center prediction error of $95.6~km$ could be achieved. $42.4~\%$ of all typhoon center predictions have absolute errors of less than $80~km$, $32.1~\%$ lie within a range of $80 - 120~km$ and the remaining $25.5~\%$ have an accuracy of above $120~km$. Three types of challenges have been revealed: the movement of a typhoon at open sea far away from land, the forecast of sudden course changes and the motion right before landfall. 

Combining observational satellite images with meteorological data from a reanalysis database has shown significant improvements. In numbers, the overall prediction error of typhoon centers could be reduced by $29.7~\%$ from $95.6~km$ to $67.2~km$. Furthermore, only $3.0~\%$ of all typhoon center predictions have an accuracy of above $120~km$. A combined $65.5~\%$ have an accuracy of less than or equal to $80~km$. It could be shown that just learning from satellite images is a good starting point but not enough for learning the complex phenomena that are responsible for the creation and motion of typhoons.

The same conclusion is made when forecasting cloud motion. Combining satellite images with physically meaningful images from reanalysis data reduces blurriness and increases the quality of predicted images. Dynamics, like the spinning motion of a typhoon, could be predicted.  

The current study is limited by the availability of satellite images. Access could only be gained to the images of those typhoons that affected the Korean peninsula, starting from $1993$. Increasing the datasets by all typhoons in East Asia could increase the prediction accuracy. However, one would need to get access to such a dataset. Another suggestion is to only use images from reanalysis data, without satellite images. The ETA interim dataset contains images of all typhoons around the world from $1979$ until today. Cloud motion could be estimated by images showing the cloud fraction. But the resolution would not be as good as the resolution of a satellite image.

\bibliography{output}

\section*{Acknowledgements}

This work was supported by the National Research Foundation of Korea (NRF) under the Project Number NRF-2017R1E1A1A03070514. Computational resources have been provided by the Flow Physics and Engineering Laboratory at Pohang University of Science and Technology.

\section*{Author contributions statement}

All authors proposed the study. M.R. processed the images and analyzed the data. S.L. provided the source code. All authors discussed the results and participated in completing the manuscript. 

\section*{Additional information}

The authors declare no competing financial interests.

\end{document}